\documentclass[fleqn,twoside]{article}
\usepackage{gc,amsfonts,bezier,epsf,cite}

\def\GR{general relativity}

\def\bh{black hole}

\def\ssph{static, spherically symmetric}

\def\mn{_{\mu\nu}}
\def\MN{^{\mu\nu}}

\def\cA{{\cal A}}

\def\od{{\overline d}}
\def\og{{\overline g}}
\def\oR{{\overline R}}

\def\r {{r_*}}

\def\M{{\mathbb M}}
\def\N{{\mathbb N}}
\def\R{{\mathbb R}}
\def\S{{\mathbb S}}

\def\ME{\mbox{$\M_{\rm E}$}}
\def\MJ{\mbox{$\M_{\rm J}$}}

\def\Str{\mbox{$\S_{\rm trans}$}}
\def\Df{\Delta\psi}

\def\Description#1{\begin{description}\itemsep -2pt
    #1 \end{description}\vspace{-5pt}}

\begin{document}
\twocolumn[
\prepno{gr-qc/0601123}{\GC {11} 305 (2005)}

\bigskip

\Title
{Generalized theories of gravity \yy and conformal continuations}

\Aunames {K.A. Bronnikov\auth{1,\dag,\ddag} and M.S. Chernakova\auth{\ddag}}

\Addresses{
\addr{\dag}
    {Centre of Gravitation and Fundamental Metrology, VNIIMS,
        46 Ozyornaya St., Moscow 119361, Russia}
\addr{\ddag}
    {Institute of Gravitation and Cosmology, Peoples' Friendship University
	    of Russia\\
	6 Miklukho-Maklaya St., Moscow 117198, Russia}
	}


\Abstract
    {Many theories of gravity admit formulations in different, conformally
     related manifolds, known as the Jordan and Einstein conformal frames.
     Among them are various scalar-tensor theories of gravity and high-order
     theories with the Lagrangian $f(R)$ where $R$ is the scalar curvature
     and $f$ is an arbitrary function. It may happen that a singularity in
     the Einstein frame corresponds to a regular surface \Str\ in the
     Jordan frame, and the solution is then continued beyond this surface.
     This phenomenon is called a conformal continuation (CC). We discuss
     the properties of vacuum static, spherically symmetric configurations
     of arbitrary dimension in scalar-tensor and $f(R)$ theories of gravity
     and indicate necessary and sufficient conditions for the existence of
     solutions admitting a CC. Two cases are distinguished, when \Str\
     is an ordinary regular sphere and when it is a Killing horizon. Two
     explicit examples of CCs are presented.  }

\bigskip
]   
\email 1 {kb20@yandex.ru}

\section{Introduction}

    High-order theories of gravity with the Lagrangian $L=f(R)$ and
    scalar-tensor theories (STT) are well-known and important alternatives to
    Einstein's \GR. They are widely used, in particular, for describing
    inflation in the early Universe \cite{inflation}, for explaining its
    present-day accelerated expansion \cite{accel} and in many other
    applications. One can also mention that curvature-nonlinear corrections
    to the Einstein theory emerge due to quantum effects of material fields
    in curved space \cite{qua}.

    There is a conformal mapping from the manifold \MJ\ with the metric
    $g\mn$, where a theory (STT or $f(R)$ gravity) is initially formulated
    (it is called the Jordan conformal frame, or Jordan picture), to the
    manifold \ME\ with the metric $\og\mn = g\mn/F(x)$ (the Einstein
    picture), in which the equations of the original theory turn into the
    equations of \GR\ with a scalar field $\phi$ endowed with a certain
    potential $V(\phi)$ (see, e.g., \cite{sok} and references therein).  If
    the conformal factor $F(x)$ is everywhere regular, then the basic
    physical properties of the manifolds \MJ\ and \ME\ coincide since, in
    such transformations, a flat asymptotic in \MJ\ maps to a flat
    asymptotic in \ME, a horizon to a horizon, a centre to a centre.
    Of special interest are, however, the cases when a singularity in \ME\
    maps (due to the properties of $F(x)$) to a regular surface in \MJ. Then
    \MJ\ may be continued in a regular manner beyond this surface; this
    phenomenon has been termed conformal continuation \cite{vac3}. In such
    cases the global properties of the manifold \MJ\ can be much richer than
    those of \ME. The new region may, in particular, contain a horizon or
    another spatial infinity.

    From a more general viewpoint, a possible existence of conformal
    continuations may mean that the observed Universe is only a region of
    a real, much larger Universe which should be described in another, more
    fundamental conformal frame. Detailed discussions of the physical
    meaning and role of different conformal frames may be found in
    Refs.\,\cite{fara,bm-erice}.

    In this work we discuss necessary and sufficient conditions for the
    existence of conformal continuations (CC) in vacuum \ssph\ space-times of
    arbitrary dimension $D\geq 3$ in STT and $f(R)$ theories and present two
    specific examples.

\section{Field equations}

\noi
{\bf $f(R)$ gravity.} Consider high-order gravity (HOG) with the action
\beq
     S_{\rm HOG} = \int d^D x \sqrt{|g|}f(R)               \label{act-hog}
\eeq
    where $f$ is a function of the scalar curvature $R$ calculated for the
    metric $g\mn$ of a space-time $\MJ = \MJ [g]$. In accord with the weak
    field limit $f\sim R$ at small $R$, we assume $f(R) >0$ and $f_R \equiv
    df/dR >0$, at least in a certain range of $R$ including $R=0$, but admit
    $f_R < 0$ and maybe $f < 0$ in general.

    The conformal mapping $\MJ \mapsto \ME$ with
\beq
    g\mn = F(\psi) \og\mn,\cm  F= |f_R|^{-2/(D-2)},
\eeq
    transforms the ``Jordan-frame'' action (\ref{act-hog}) into the
    Einstein-frame action
\beq                                   			     \label{act-E}
    S = \int d^D x \,\sqrt{|\og|} [\oR + (\d\psi)^2 - 2V(\psi)],
\eeq
    where
\bear
    \psi \eql \pm \sqrt{\frac{D-1}{D-2}}\log |f_R|,
\nn            						\label{phi-hog}
    2V(\psi) \eql |f_R|^{-D/(D-2)} (R|f_R| - f).
\ear
    The field equations due to (\ref{act-hog}) after this substitution
    turn into the field equations due to (\ref{act-E}). Let us write them
    down for \ssph\ configurations, taking the metric $\og\mn$ in the form
\beq \nq\,                                                       \label{ds_E}
    ds_E^2 = \og\mn dx^\mu dx^\nu =
        A(\rho) dt^2 - \frac{d\rho^2}{A(\rho)} - r^2(\rho) d\Omega_\od{}^2,
\eeq
    where $d\Omega_\od{}^2$ is the linear element on a sphere $\S^{\od}$
    of unit radius, and $\psi=\psi(\rho)$. Three independent combinations of
    the Einstein equations can be written as
\bear
      (A_\rho r^\od)_\rho \eql - (4/\od)r^\od V,                \label{00E}
\\
    \od r_{\rho\rho}/r  \eql -{\psi_\rho}^2,                    \label{01E}
\\
    A (r^2)_{\rho\rho} - r^2 A_{\rho\rho}\lal
\nn
     + (\od-2)r_\rho(2Ar_\rho - A_\rho r)\eql 2(\od-1);
                                                		\label{02E}
\ear
    where the subscript $\rho$ denotes $d/d\rho$. The scalar field equation
    follows from the Einstein equations. Given a potential $V(\psi)$,
    (\ref{00E})--(\ref{02E}) is a determined set of equations for the
    unknowns $r,\ A,\ \psi$.

    The metric $g\mn = F\og\mn$ may be taken in a form similar to
    (\ref{ds_E}):
\beq                         \nq                               \label{ds_J}
    ds_J^2 = g\mn dx^\mu dx^\nu =
        \cA(q) dt^2 - \frac{dq^2}{\cA(q)} - \r^2(q) d\Omega_\od{}^2.
\eeq
    The quantities in (\ref{ds_J}) and (\ref{ds_E}) are related by
\beq            \nq
              \cA(q) = F A(\rho),                               \label{to_q}
     \quad  \r^2 (q) = Fr^2 (\rho),
     \quad        dq = \pm F d\rho.
\eeq


    In both metrics (\ref{ds_E}) and (\ref{ds_J}) we have chosen the
    ``quasiglobal'' radial coordinates \cite{vac1} ($\rho$ ¨ $q$,
    respectively), which are convenient for describing Killing horizons:
    near a horizon $\rho=\rho_h$, the function $A(\rho)$ behaves as
    $(\rho-\rho_h)^k$ where $k$ is the horizon order: $k=1$ corresponds to
    a simple, Schwarzschild-type horizon, $k=2$ to a double horizon, like
    that in an extremal Reissner-Nordstr\"om \bh\ etc. The function $\cA(q)$
    plays a similar role in the metric (\ref{ds_J}).

\medskip\noi
{\bf General scalar-tensor theory.} The action in \MJ, instead of
    (\ref{act-hog}), has the form
\bearr  \nq
     S_{\rm STT} = \int \! d^D x \sqrt{|g|}                \label{act-stt}
           [f(\phi) R + h(\phi) (\d\phi)^2 -2 U(\phi)],
\ear
    where $f$, $h$ and $U$ are functions of the real scalar field
    $\phi$, $(\d\phi)^2 = g\MN\d_\mu\phi\d_\nu\phi$.

    The conformal mapping $\MJ \mapsto \ME$ with
\beq
	g\mn = F(\psi) \og\mn,\cm  F= |f|^{-2/(D-2)},        \label{wag}
\eeq
    transforms (\ref{act-stt}) into the same Einstein-frame action
    (\ref{act-E}), where
\bearr
        \frac{d\psi}{d\phi} = \pm                          \label{phi-wag}
                \frac{\sqrt{|l(\phi)}|}{f(\phi)},\cm
    l(\phi) \eqdef fh + \frac{D{-}1}{D{-}2}\Bigl(\frac{df}{d\phi}\Bigr)^2,
\nnn
                   V(\psi) = |f|^{-D/(D-2)} (\phi)\, U(\phi).
\ear
    In \ME\ we have the same \eqs (\ref{00E})-(\ref{02E}).

\section{Conformal continuations: conditions and properties}

    Consider the possible situation when the metric $\og\mn$ is singular at
    some value of $\rho$ while $g\mn$ at the corresponding value of $q$ is
    regular. In such a case $\MJ$ can be continued in a regular manner
    through this surface (to be denoted $\Str$), i.e., by definition
    \cite{vac3,vac4}, we have a conformal continuation (CC).

    In our case of spherical symmetry, the sphere $\Str \in \MJ$ may be
    either an ordinary sphere, where both metric coefficients $\r^2$ and
    $\cA$ are finite (we label such a continuation CC-I), or a Killing
    horizon at which $\r^2$ is finite but $\cA =0$ (to be labelled CC-II).

    Without loss of generality, we suppose for convenience that at $\Str$
    the coordinate values are $\rho = 0$ and $q = 0$ and $\rho > 0$
    in $\ME$ outside $\Str$. According to
    (\ref{to_q}), we must have, in terms of $\og\mn$,
\beq
     F^{-1}\sim r^2 \to 0                               	\label{r0}
                \cm {\rm as} \qquad \rho\to 0,
\eeq
    and, in addition, $A(\rho) \sim r^2(\rho)$ for CC-I, whereas for
    CC-II we must have in $\MJ$: $\cA (q) \sim q^n$ at small $q$, where
    $n\in \N$ is the order of the horizon.

    Let us use the field equations in $\ME$ for further estimates.
    \eq (\ref{02E}) may be put in the form
\beq                                                        \label{02E'}
    \frac{d}{d\rho}\biggl(r^D \frac{dB}{d\rho}\biggr) = -2(\od-1)r^{\od-2},
\eeq
    where the function $B(\rho) = A/r^2 = B(q) = \cA/\r^2$ is invariant
    under conformal transformations and should be finite at $\rho = q = 0$.
    Moreover, since $\Str$ is a regular sphere in $\MJ$, $B(q)$ should be a
    smooth function near $q=0$.

    It can be shown \cite{hog1} that: 1) for $D=3$ a CC can only exist if $B
    = B_0 =\const$; 2) for $D > 3$ the function $B(q)$ behaves near $\Str$
    as
\beq
        B(q) = B_0 + \half B_2 q^2 + o(q^2), \cm B_2 \ne 0,      \label{B_}
\eeq
    where $B_2 < 0$, i.e., the function $B(q)$ has a maximum at $q=0$.

    All this was obtained by comparing the metrics $g\mn$ and $\og\mn$,
    {\sl without specifying a theory in which the CC takes place, and for
    both kinds of transitions, CC-I and CC-II. Both kinds of transitions are
    thus possible for $D > 3$, and, in particular, in CC-II $\Str$ is a
    double horizon connecting two T regions} (since $B = \cA/\r^2$ is
    negative at both sides of $\Str$).

    In 3D gravity only CC-I can take place: a horizon, at which $B=0$ but
    $B\ne 0$ in its neighbourhood, is inconsistent with the condition $B =
    \const$.

\subsection{Conformal continuations in $f(R)$ theories}

    Now, for the theory (\ref{act-hog}), the CC conditions can be made more
    precise. A transition surface $\Str$ should correspond to values of $R$
    at which the function $F(\psi)$ tends to infinity, i.e., where $f_R =0$.
    In this case, according to (\ref{r0}), near $\rho=0$ we have
    $f_R^{-2/\od} \sim r^{-2}$. Using the expression for $\psi$ in
    (\ref{phi-hog}) and \eq (\ref{01E}), we obtain:  $r \approx \const\cdot
    \rho^{1/D}$ as $\rho\to 0$. According to (\ref{to_q}), we also obtain:
    $F \sim \rho^{-2/D}$ and the relation $q\sim \rho^{1-2/D}$ between
    the coordinates $\rho$ and $q$ at their small values. The results can be
    summarized in the following theorem \cite{hog1}:

\Theorem{Theorem 1}
    {For a \ssph\ configuration in the theory
    (\ref{act-hog}) in $D\geq 3$ dimensions  the following
     necessary conditions and properties of
    a CC (at $\rho=q=0$) take place:
\Description{
\item[(a)]
	$f(R)$ has an extremum, at which $f_R = 0$ and $f_{RR}\ne 0$;
\item[(b)]
	$dR/dq \ne 0$ at $q=0$, hence the ranges of the curvature $R$ are
        different at the two sides of $\Str$;
\item[(c)]
	in the Einstein frame, $r(\rho) \sim \rho^{1/D}$ as $\rho\to 0$;
\item[(d)]
	in the Jordan frame, $B(q)$ has a maximum at $q=0$.
\item[(e)]
	For $D=3$, $B(\rho) = B(q) =\const$.
\item[(f)]
        A CC-II is only possible for $D\geq 4$, and $\Str$ is then a double
    	horizon connecting two T regions.
    }}

     One can prove \cite{hog1} that the above necessary conditions are also
     sufficient for the existence of a CC. This is done by seeking
     the unknown functions in the field equations in \MJ\ in the form
     of Taylor expansions in $q$. Evidently, CC-I are of more general nature
     than CC-II since the existence of a double horizon is a very special
     condition for the metric, expressed in the initial condition $B(0) = 0$
     in the field equations.

\medskip\noi
{\bf Example.} Consider an example of an exact solution to the field
     equations with CC-I in $D=4$-dimensional space-time. In the Jordan
     frame \MJ\ it is given by the functions
\bearr
     f = -ac R + 2c\sqrt{R} = 2c/q-ac/q^2, \qquad     \r = q,
\nnn
	B = (3q-2a)/6q^3, \qquad R = 1/q^2,
\ear
    where $a,\ c - \const > 0$. We take for convenience $a=1$, $c=1$
    (choosing the appropriate units). Then $f_R = 0$ at $q =q_{\rm trans}=1$.

    The Jordan and Einstein metrics are
\bear
    ds_J^2 \eql \left(\frac{1}{2}-\frac{1}{3q}\right)dt^2-
        \left(\frac{1}{2}-\frac{1}{3q}\right)^{-1}dq^2-q^{2}d\Omega^2,
\nn
     ds_E^2 \eql |q-1| ds_J^2.
\ear
    Thus the Jordan-frame metric has a form close to Schwarz\-schild's, it is
    singular at the centre $q=0$ and has a horizon at $q = 2/3$. Its
    asymptotic is non-flat due to a solid angle deficit equal to $2\pi$,
    i.e., it has the same nature as the asymptotic of a global monopole (as
    can be easily seen by changing the coordinates from $t$ and $q$ to
    $\bar t = t/\sqrt{2}$ and $\bar q = q\sqrt{2}$). In \ME, the metric is
    singular at $q=0$ and $q=1$ and contains a horizon at $q=2/3$. The
    manifold $\MJ$ has two Einstein couterparts $\ME_1$ and $\ME_2$,
    existing separately for $q > 1$ and $q < 1$. The first of them has a
    non-flat asymptotic as $q \to \infty$ and a naked singularity at the
    centre ($q=1$), the other has two singular centres at $q=0$ and $q=1$,
    separated by a horizon at $q=2/3$.

    The scalar field and its potential in \ME\ have the form
\[
    \psi = \pm \sqrt{3/2}\ln|q-1|, \cm V = -\half q^{-1}(q-1)^{-2}.
\]

    This example is of methodological nature and demonstrates an
    essential distinction between the descriptions of the theory in the
    Jordan and Einstein pictures.

\subsection{Conformal continuations in STT}

    Assume now that there is a STT (\ref{act-stt}) given in \MJ. In this case
    a CC from \ME\ into \MJ\ can occur at such values of the scalar field
    $\phi$ that the conformal factor $F$ is singular while the functions
    $f$, $h$ and $U$ in the action (\ref{act-stt}) are regular.  This means
    that at $\phi=\phi_0$, corresponding to a possible transition surface
    \Str, the function $f(\phi)$ has a zero of a certain order $n$. We then
    have in the transformation (\ref{phi-wag}) near $\phi=\phi_0$ in the
    leading order of magnitude
\beq
    f(\phi) \sim \Df^n, \quad   n = 1,2,\ldots,
         \qquad        \Df \equiv  \phi-\phi_0.      \label{phi0}
\eeq
    One can notice, however, that $n > 1$ leads to $l(\phi_0) = 0$
    (recall that by our convention $h(\phi)\equiv 1$). This generically
    leads to a curvature singularity in \MJ, and though such a singularity
    can be avoided at some special choices of $f$ and $U$, we will ignore
    this possibility and simply assume $l > 0$ at \Str.

    Thus, according to (\ref{phi-wag}), near \Str\ ($\phi=\phi_0$)
\bear
    f (\phi) \sim \Df \sim \e^{-\psi\sqrt{\od/(\od+1)}},       \label{Df}
\ear
    where without loss of generality we choose the sign of $\psi$ so that
    $\psi\to\infty$ as $\Df \to 0$.

    One can deduce from (\ref{Df}), (\ref{r0}) and (\ref{01E}) that near
    \Str\ it holds $r(\rho) \sim \rho^{1/D}$. It follows that both $\Df$ and
    $q$ behave as $r^\od$ in the neighbourhood of \Str\, hence $d\psi/dq$ is
    finite.

\medskip\noi
{\bf 1. \ Continuation through an ordinary sphere (CC-I).}  A CC-I can occur
    if $F(\psi) = |f|^{-2/\od} \sim 1/r^2 \sim 1/A$ as $\psi\to\infty$,
    while the behaviour of $f$ is specified by (\ref{Df}).  The following
    theorem is valid \cite{vac4}:

\Theorem{Theorem 2}
    {Consider scalar-vacuum configurations with the metric (\ref{ds_J})
    and $\phi=\phi(q)$ in the theory (\ref{act-stt}) with $h(\phi) \equiv 1$
    and $l(\phi) >0$. Suppose that $f(\phi)$ has a simple zero at some
    $\phi = \phi_0$, and $|U(\phi_0)| < \infty$. Then:
\Description{
\item[(i)]
    there exists a solution in \MJ, smooth in a neighbourhood of the surface
    \Str\ ($\phi=\phi_0$), which is an ordinary regular surface in \MJ;
\item[(ii)]
    in this solution the ranges of $\phi$ are different at different sides
    of \Str.  }}

    Thus, in fact, the CC necessary conditions turn out to be sufficient.

\medskip\noi
 {\bf 2. Continuation through a horizon in \MJ\ (CC-II).}
    In this case we have near $q=0$:
    $f(\phi) \sim \Df$, $\cA(q) = AF \sim q^k$, $\r^2(q) = Fr^2 = O(1)$.
    The following theorem describes the necessary and sufficient
    conditions for the existence of CC-II \cite{vac4}:

\Theorem{Theorem 3}
    {Consider scalar-vacuum configurations with the metric (\ref{ds_J})
    and $\phi=\phi(q)$ in the theory (\ref{act-stt}) with $h(\phi) \equiv 1$
    and $l(\phi) >0$. Suppose that $f(\phi)$ has a simple zero at some
    $\phi = \phi_0$.
    There exists a solution in \MJ, smooth in a neighbourhood of the
    surface \Str\ ($\phi=\phi_0$), which is a Killing horizon in \MJ,
    {\bf if and only if:}
\Description{
\item[(a)]  $D\geq 4$,
\item[(b)]  $\phi_0$ is a simple zero of $U(\phi)$,
\item[(c)]  $dU/df > 0$ at $\phi=\phi_0$.
    }
    Then, in addition,
\Description{
\item[(d)]
     \Str\ is a second-order horizon, connecting two T-regions in \MJ;
\item[(e)]
     the ranges of $\phi$ are different at different sides of \Str.
    }}

    Thus the only kind of STT configurations admitting CC-II is a $D\geq
    4$ Kantowski-Sachs cosmology consisting of two T-regions (in fact,
    epochs, since $\rho$ is a temporal coordinate), separated by a
    second-order horizon.

\medskip\noi
{\bf Example: Conformal scalar field in GR.}
    Consider an explicit example of configurations with CC-I for $D=4$.
    This example is well known \cite{br73} and is given here to illustrate
    the generic character of wormholes appearing due to CC.

    The conformal scalar field in GR can be viewed as a special case of STT,
    such that in (\ref{act-stt})
\beq                                                         \label{conf}
    f(\phi) = 1 - \phi^2/6, \qquad  h(\phi)=1, \qquad U(\phi) =0.
\eeq
    After the transformation $g\mn = F(\phi)\og\mn$ with
\bearr
       \phi = \sqrt{6} \tanh (\psi+\psi_0)/\sqrt{6}),
\nnn
	F(\phi) = \cosh^2[(\psi+\psi_0)/\sqrt{6}], \cm \psi_0=\const,
                        \label{phi-conf}
\ear
    we obtain the action (\ref{act-E}) with $V\equiv 0$. The corresponding
    \ssph\ solution is well known: it is Fisher's solution \cite{fisher}. In
    terms of the harmonic radial coordinate $u\in \R_+$, specified by the
    condition $g_{uu}=-g_{tt}(g_{\theta\theta})^2$, the solution is
    \cite{br73}
\bearr
    ds^2_{\rm E} = \e^{-2mu}dt^2                       	     \label{fish1}
                    - \frac{k^2\e^{2mu}}{\sinh^2(ku)}
        \biggl[\frac{k^2 du^2}{\sinh^2(ku)} + d\Omega^2\biggr],
\nnn
	\psi = C u,
\ear
    where $m$ (mass), $C$ (scalar charge), $k>0$ and $u_0$ are integration
    constants, and $k$ is expressed in terms of $m$ and $C$:
    $k^2 = m^2 + C^2/2$.

    Another convenient form of the solution is obtained in isotropic
    coordinates: with $y = \tanh (ku/2)$, \eqs (\ref{fish1}) are converted to
\bearr
    ds^2_{\rm E} =                                    	     \label{fish2}
            A(y)\,dt^2
    	        - \frac{k^2(1-y^2)^2}{y^4 A(y)} (dy^2 + y^2 d\Omega^2),
\nnn
     A(y) = \biggl|\frac{1-y}{1+y}\biggr|^{2m/k},
\cm
     \psi =\frac{C}{k}\ln \biggl|\frac{1+y}{1-y}\biggr|.
\ear

    The solution is asymptotically flat at $u\to 0$ ($y \to 0$), has
    no horizon when $C\ne 0$ and is singular at the centre ($u\to \infty$,
    $y\to 1-0$, $\psi\to\infty$).

    The Jordan-frame solution for (\ref{conf}) is described by the metric
    $ds^2 = F(\psi) ds_{\rm E}^2$ and the $\phi$ field according to
    (\ref{phi-conf}). It is the conformal scalar field solution \cite{bbm70,
    bek74}, its properties are more diverse and can be described as
    follows (putting, for definiteness, $m > 0$ and $C > 0$):

\medskip\noi
{\bf 1.} $C < \sqrt{6}m$. The metric behaves qualitatively as in Fisher's
    solution: it is flat at $y\to 0$ ($u\to 0$), and both $g_{tt}$ and
    $r^2=|g_{\theta\theta}|$ vanish at $y = 1$ ($u\to\infty$) --- a singular
    attracting centre. A difference is that here the scalar field is finite:
    $\phi \to \sqrt{6}$.

\medskip\noi
{\bf 2.} $C > \sqrt{6}m$. Instead of a singular centre, at $y\to 1$
    ($u \to \infty$) one has a singularity of infinite radius:
    $g_{tt}\to \infty$ and $r^2 \to \infty$. Again $\phi\to \sqrt{6}$.

\medskip\noi
{\bf 3.} $C = \sqrt{6}m$, $k=2m$. Now the metric and $\phi$ are
    regular at $y = 1$; it is \Str, and the coordinate $y$ provides a
    continuation. The solution acquires the form
\bear                    \nq
     ds^2 \eql \frac{(1{+}yy_0)^2}{1-y_0^2}\Biggl[\frac{dt^2}{(1{+}y)^2}
               -\frac{m^2(1{+}y)^2}{y^4}(dy^2{+}y^2 d\Omega^2)\Biggr],
\nn
     \phi \eql \sqrt{6} \frac{y+y_0}{1 + yy_0},          \label{con-y}
\ear
    where $y_0 = \tanh (\psi_0/\sqrt{6})$. The range $u\in \R_+$,
    describing the whole manifold \ME\ in Fisher's solution, corresponds
    to the range $0 < y < 1$, describing only a region $\MJ'$ of the
    manifold \MJ\ of the solution (\ref{con-y}). The properties of the
    latter depend on the sign of $y_0$ \cite{br73}. In all cases, $y=0$
    corresponds to a flat asymptotic, where $\phi \to \sqrt{6}y_0$, $|y_0|
    < 1$.

\medskip\noi
{\bf 3a:} $y_0 < 0$. The solution is defined in the range $0 < y <1/|y_0|$.
    At $y=1/|y_0|$, there is a naked attracting central singularity:
    $g_{tt}\to 0$, $r^2\to 0$, $\phi\to\infty$.

\medskip\noi
{\bf 3b:} $y_0 > 0$. The solution is defined in the range $y\in \R_+$.
    At $y\to\infty$, we find another flat spatial infinity, where
    $\phi\to \sqrt{6}/y_0$, $r^2\to\infty$ and $g_{tt}$ tends to a finite
    limit. This is a {\sl wormhole solution\/} found in Ref.\,\cite{br73}
    and recently discussed in Ref.\,\cite{barvis}.

\medskip\noi
{\bf 3c:} $y_0=0$, $\psi=\sqrt{6}y$, $y\in \R_+$. In this case it is helpful
    to pass to the conventional coordinate $r$: $y=m/(r-m)$. The solution
    is the well-known black hole with a conformal scalar field
    \cite{bbm70,bek74}.

    The whole manifold \MJ\ can be represented as the union
    $\MJ = \MJ' \cup \Str \cup \MJ''$ where $\MJ'$ is the region $y<1$,
    which is, according to (\ref{phi-conf}), in one-to-one correspondence
    with the manifold \ME\ of the Fisher solution (\ref{fish1}). The
    ``antigravitational'' ($f(\psi) < 0$) region $\MJ''$ ($y > 1$) is in
    similar correspondence with another ``copy'' of the Fisher solution,
    where, instead of (\ref{phi-conf}),
\beq
      \phi = \sqrt{6} \coth (\psi/\sqrt{6}),  \qquad \label{phi-con'}
      F(\psi) = \sinh^2(\psi/\sqrt{6}).
\eeq

\section {Concluding remarks}

    Studying \ssph\ vacuum solutions in two important classes of theories of
    gravity, $f(R)$ theories and STT, we have demonstrated that conformal
    continuations (CC) are quite a widespread phenomenon. It has also been
    shown \cite{vac4} that one of generic types of configurations in the
    Jordan picture in STT, existing due to CC, are traversable wormholes. In
    $f(R)$ theory, one can also expect the existence of non-singular vacuum
    solutions of physical interest.

    The results presented here may be extended to electrovacuum
    solutions of the same classes of theories as well as to solutions of a
    more general class of theories, unifying these two, with the action
\beq
      S = \int d^D x \sqrt{|g|}\,\biggl[ f(R,\phi) \pm (\d\phi)^2 \biggr],
\eeq
    where $f$ is an arbitrary (sufficiently smooth) function of two
    variables. Such theories also admit a transition to the Einstein picture,
    but with two scalar fields combined to a kind of sigma model \cite{fara}.
    Different conformal relations between STT and $f(R)$ theories are also
    discussed in Ref.\,\cite{od05}

    It has been shown that many wormhole solutions with CC in STT are
    unstable under monopole perturbations, but the instability weakens with
    growing electric charge \cite{instab}. Many questions are yet to be
    answered before one could judge whether or not configurations with CC
    can describe the space-times able to exist in nature.

\end{document}